\documentclass[twocolumn,nofootinbib,showpacs,prd]{revtex4}

\usepackage{graphics,graphicx}
\usepackage{amsmath}
\usepackage{psfrag}
\usepackage{dcolumn}
\usepackage{psfrag}

\def\no{\nonumber \\ & \quad}
\def\noQ{\nonumber \\}
\def\bes{\begin{subequations}}
\def\ens{\end{subequations}}
\def\bal{\begin{align}}
\def\eal{\end{align}}

\newcommand{\vek}[1]{\boldsymbol{#1}}

\begin{document}

\title{
Binary Black Hole Coalescence in Semi-Analytic Puncture 
Evolution 
}

\author{ Achamveedu Gopakumar and Gerhard Sch\"afer
}
\affiliation{Theoretisch-Physikalisches Institut, Friedrich-Schiller-Universit\"at Jena,
Max-Wien-Platz 1, 07743 Jena, Germany}

\date{\today}

\begin{abstract}
Binary black-hole coalescence is treated semi-analytically
by a novel approach.
Our prescription employs
the conservative Skeleton Hamiltonian
that describes orbiting
Brill-Lindquist wormholes (termed punctures in Numerical
Relativity) within a waveless truncation to the
Einstein field equations [G.~Faye, P.~Jaranowski and G.~Sch\"afer,
Phys.\ Rev.\  D {\bf 69}, 124029 (2004)].
We incorporate, in a transparent Hamiltonian way and in
Burke-Thorne gauge structure, the
effects of gravitational radiation reaction into the
above Skeleton dynamics with the help of 3.5PN accurate
angular momentum flux for compact binaries in quasi-circular orbits
to obtain a
Semi-Analytic Puncture Evolution to model merging black-hole binaries.
With the help of the TaylorT4 approximant at 3.5PN order, we 
perform a {\it first-order} comparison between 
gravitational wave phase evolutions in Numerical Relativity 
and our approach for equal-mass binary black holes.
This comparison reveals that a modified Skeletonian reactive dynamics
that employs flexible parameters will be required to 
prevent the dephasing between our scheme and Numerical Relativity,
similar to what is pursued in the Effective One Body approach.
A rough estimate for the gravitational waveform associated with
the binary black-hole coalescence in our approach is also provided.

\end{abstract}

\pacs{
04.30.Db, 
04.25.Nx 
}

\maketitle

\section{Introduction}

 Binary black hole mergers are the most plausible sources of
gravitational radiation for the 
currently operational and planned ground-based laser interferometric
gravitational-wave (GW) detectors like GEO-LIGO and VIRGO.
Further, coalescing binary black holes (BBH) are considered to be the workhorse sources to
realize GW astronomy with the proposed 
space-based GW interferometer LISA.
Recent advances in Numerical Relativity, achieved by several independent groups, should enable 
its practitioners to create  accurate and `exact' numerical relativity based GW search templates
for the late stages of binary black hole coalescence \cite{NR_Golm}.
However, it is also desirable to develop semi-analytic prescriptions,
based on well defined assumptions, capable of 
describing the most relativistic part of the binary black hole coalescence, namely
the last few cycles of the inspiral, plunge and the subsequent merger.
The two main reasons to develop such descriptions are as follows.
It is reasonable to imagine that
semi-analytic approaches, based on certain assumptions, should be able
to probe the purely numerical approaches in order to extract
the `physics' behind binary black hole coalescence.
The second reason is related to the GW data analysis requirement of creating 
huge banks of accurate GW templates to cover, rather densely, the binary black hole
parameter space.

  In the literature and to best of our knowledge, there exists only two approaches that can
semi-analytically model the late stages of non-spinning compact binary coalescence.
Historically,
the first prescription is the so-called hybrid approach, developed by Will and his 
collaborators, which re-sums exactly the `Schwarzschild terms' 
in the post-Newtonian(PN) accurate equations of motion for compact binaries and treats 
the symmetric mass ratio terms  as additional 
corrections \cite{KWW}. The second and widely employed approach is the so-called 
Effective One-Body (EOB) scheme, developed by Damour and 
collaborators \cite{EOB,EOB_Golm}.
The EOB scheme requires available  PN accurate results for the conservative orbital 
dynamics and GW luminosity associated with comparable mass compact binaries 
moving along quasi-circular orbits 
extractable  
from Refs.~\cite{3PN_res}. In the EOB approach, one packages these PN
accurate results in a certain re-summed form with the hope of extending 
the validity of PN accurate results to strong-field (and fast-motion) regimes.
The EOB package allows PN accurate  dynamics of a comparable mass compact 
binary, having masses $m_1$ and $m_2$, to be mapped onto an
essentially geodesic dynamics of a test-particle of mass $\mu =m_1\,m_2/(m_1+m_2)$ moving
in an $\eta= m_1\,m_2/(m_1+m_2)^2$ deformed Schwarzschild  geometry. 
Further, reactive dynamics is included with the help of certain 
Pad\'e re-summed radiation reaction force.
It is important to note that in the hybrid and EOB approaches, describing 
late stages of binary black hole coalescence,  essentially contain a   
single black hole and some undesirable features of these schemes were 
pointed out in Refs.~\cite{WS} and \cite{DG06}.

  In this paper, we present a prescription that can semi-analytically model the late
stages of coalescing non-spinning binary black holes (BBH).
Our approach employs the `Skeleton' solution to the Einstein field equations,
detailed in Ref.~\cite{FJS}, 
to describe the BBH conservative dynamics.
We augment the above mentioned BBH conservative dynamics by
incorporating radiation reaction effects in a fully Hamiltonian way 
 with the help of the 3.5PN accurate GW luminosity, available in Refs.~\cite{3PN_res}.
This is how, in a nutshell, we provide a semi-analytical description for the late stages of BBH 
coalescence. The fact that the Skeleton Hamiltonian is constructed in the 
Arnowitt-Deser-Misner (ADM) 
$3+1$ approach to General Relativity \cite{ADM} makes our scheme very close to the BBH dynamics,  
analyzed by several Numerical Relativity groups. 
The Skeleton Hamiltonian contains, in its static part, the potential
of the well-known Brill-Lindquist BBH initial value configuration \cite{BL,JS99}.
Further, it allows black holes to orbit each other along conservative orbital 
configurations. 
We emphasize that the above mentioned conservative orbital evolution obey exactly 
only a truncated version of  Einstein's field
equations \cite{FJS}.  
In the Skeleton approach for black holes in circular orbits,  
Ref.~\cite{FJS} also demonstrated 
an intriguing property; the ADM mass (identified through the 3-metric) \cite{ADM}
coincides with the Komar mass (identified through the pure time
component of the 4-metric)\cite{Kommar}.
We note that the equality of ADM and Komar masses is 
a general property of stationary
solutions to the Einstein field equations for isolated systems. 
This is, in our view, an appealing feature of the Skeleton scheme.

In Numerical Relativity simulations that
employ punctures to model black holes, it is customary to
use  Brill-Lindquist configurations as initial data \cite{BB}.
Therefore, the unique construction of the Skeleton Hamiltonian for BBH, based 
on certain well defined assumptions, and its present extension   
should allow us to probe the various aspects of Numerical Relativity based BBH simulations 
involving punctures. This prompted us to name
our approach as Semi-Analytic Puncture Evolution (SAPE).
However, it should be noted that the recent Numerical Relativity
simulations involving black holes do not impose
conformal flat condition in their evolutions, required by our SAPE.

       The organization of this paper is as follows. In the next section, we 
briefly summarize the Skeleton approximation and the 
conservative binary black hole dynamics it defines, detailed in Ref.~\cite{FJS}. 
Section~\ref{sec:Skdy} also includes a fully consistent Hamiltonian
way of incorporating  
the effects of radiation reaction on to the conservative Skeleton dynamics. 
The resulting binary black hole dynamics and its implications are explored 
in Section~\ref{Skdyn_results}.
An estimate for the complete gravitational wave polarizations 
associated with the binary black hole coalescence in reactive Skeleton dynamics, 
in the `restricted waveform' approach,
is presented in Section~\ref{Skdyn_h}.
Our conclusions and future directions are  available in Section~\ref{Skdyn_CR}. 

\section{ 
Binary Puncture Dynamics in Skeleton Approximation }
\label{sec:Skdy}

In this section, we briefly summarize the construction of the conservative binary black hole Skeleton Hamiltonian,
detailed in Ref.~\cite{FJS}.
Afterwards, we explain our prescription for 
 including the effects of GW radiation reaction into the 
Skeletonian conservative dynamics.

\subsection{Conservative Skeletonian Binary Black Hole Dynamics}




In the ADM formulation of General Relativity, the spacetime line element
in the $(3+1)$ decomposed form is given by
\begin{align}
ds^2 &= - \alpha^2 c^2 dt^2 + \gamma_{ij}(dx^i + \beta^icdt)(dx^j+\beta^jcdt)
\no
\mbox{ ( $i,j = 1,2,3 $ ) },
\end{align}
where $\alpha$ is the lapse function, $\beta^i$ the shift vector,
$\gamma_{ij}$ the induced metric on a three-dimensional
spatial slice $\Sigma (t)$, parametrized by the
time coordinate $t$, and $c$ is the velocity of
light. 
The canonical conjugate to the 3-metric $\gamma_{ij}$ is given by
$\pi^{ij}c^3/16\pi G$, where $G$ is the Newtonian gravitational constant and
it is a tensor  density of weight $+1$. 
Further, $\gamma_{ij} $ and $\pi^{ij} $ satisfy  the Hamiltonian and  momentum constraints
\cite{ADM}. 
On any three-dimensional
spatial slice $\Sigma (t)$, the above four constraints may be written as 
\begin{subequations}
\label{Eq2_Sk}
\begin{align}
\gamma^{1/2}\mbox{R} - \frac{1}{\gamma^{1/2}}\left(\pi^i_j \pi^j_i-\frac{1}{2}\pi^2\right)
&=
\frac{16\pi G}{c^3} 
\sum_a\, \biggl ( m_a^2c^2 
\no
+ \gamma^{ij}p_{ai}p_{aj} \biggr )^{1/2}\delta_a \,,
\\
-2 \partial_j \pi^j_i + \pi^{kl} \partial_i \gamma_{kl}  &= \frac{16\pi G}{c^3} \sum_a p_{ai}\delta_a\,,
\end{align}
\end{subequations}
where 
$\mbox{R}$, $\gamma$, $\gamma^{ij}$ stand respectively for the curvature scalar,
the determinant, the inverse metric associated with the 3-metric $\gamma_{ij}$.
The linear momentum of point-mass $a$
($a$=1,2) is denoted by $p_{ai}$ and its bare mass by $m_a$.
The partial derivative with respect to the hypersurface coordinates
is denoted by $\partial_i = \partial / \partial x^i$.
The Dirac delta function $\delta_a =
\delta(x^i-x^i_a(t))$ is defined by $\delta (x^i-x^i_a(t)) = 0$ if $x^i \ne x^i_a(t)$
and $\int d^3x~ \delta (x^i-x^i_a(t))
=1$, where $x^i_a(t)$ is the space coordinate of point-mass $a$ at time $t$.

  The ADM coordinate conditions that generalize the isotropic
Schwarzschild metric read
\begin{subequations}
\label{Eq_Sk_3}
\begin{align}
\label{Eq_Sk_3a}
\gamma_{ij}&=\left(1+\frac{1}{8}\phi\right)^4\delta_{ij}+h^{\rm TT}_{ ij}\,,\\
\label{Eq_Sk_3b}
\pi^{ii}&=0\,,
\end{align}
\end{subequations}
where $\phi$ is a function vanishing at spatial infinity. 
The tranverse-traceless (TT) part of the
metric $\gamma_{ij}$ with respect to the Euclidean 3-metric $\delta_{ij}$ is denoted by
$h^{TT}_{ij}$
and the Einstein summation convention is also employed in the above equation, $\pi^{ii}=0$.
Taking into account the above gauge condition for $\pi^{ij}$,
the following decomposition can be achieved:
\begin{align}
\pi^{ij}={\tilde{\pi}^{ij}}+\pi_{\rm TT}^{ij}\,,
\end{align}
where ${\tilde{\pi}^{ij}}$ denotes the longitudinal part of $\pi^{ij}$ which may 
be expressed as
\begin{align}
\label{piV_def}
\tilde{\pi}^{ij}=\partial_i \pi^j + \partial_j \pi^i -
\frac{2}{3}\delta_{ij} \partial_k \pi^k\,,
\end{align}
where $\pi^i$ is a vector in the flat 3-space.
Further, the TT part of $\pi^{ij}$, or rather $\pi_{\rm TT}^{ij}c^3/16\pi G$, provides the canonical conjugate
to $h^{\rm TT}_{ij}$.

The Hamilton functional for non-spinning binary black hole
system is given by
\begin{align}
H\left[x^i_a, p_{ai}, h^{\rm TT}_{ij}, \pi_{\rm TT}^{ij}\right]
&= - \frac{c^4}{16\pi G}\int d^3x~
\Delta\phi\,\biggl [ x^i_a, p_{ai}, 
\no
h^{\rm TT}_{ij}, 
\pi_{\rm TT}^{ij}
\biggr ]\,,
\end{align}
where $\Delta$ is the Laplacian in the $3$-dimensional flat space.
While dealing with the dynamics of 
binary black-hole systems, it is highly advantageous
to introduce a Routh functional of the form,
\begin{align}
R\left[x^i_a, p_{ai}, h^{\rm TT}_{ij}, \partial_th^{\rm TT}_{ij}\right]
&= H 
- \frac{c^3}{16\pi G} \int d^3x ~\pi^{ij}_{\rm TT} \partial_th^{\rm TT}_{ij}\,.
\end{align}
This functional is a Hamiltonian for the particle (point-mass) degrees of freedom,
and a Lagrangian for the independent gravitational field degrees of freedom.
The truncation implemented in Ref.~\cite{FJS}, {\it i.e.} to put $h^{\rm TT}_{ij}
\equiv 0$, is called the conformal flat condition and it results in 
$H=R(x^i_a, p_{ai})$.

   It is possible to show that the introduced source model, namely $\delta_a$ terms in Eqs.~(\ref{Eq2_Sk}),
can  produce certain binary black-hole spacetime.
Indeed, it was demonstrated in Refs.~\cite{JS99} that one  
obtains the Brill-Lindquist initial value
solution for uncharged black holes at rest, available in Ref.~\cite{BL}.
Therefore, let us first
summarize Refs.~~\cite{JS99}.
In Refs.~\cite{JS99}, one imposes $h^{\rm TT}_{ij} =0$ and
$p_{ai}=0$ (and this leads to $\pi^{ij} =0$). Under these restrictions,
the Hamiltonian constraint becomes  
\begin{eqnarray}
- \left(1+\frac{1}{8}~\phi\right) \Delta \phi =
\frac{16\pi G}{c^2} \sum_a m_a \delta_a.
\end{eqnarray}
The exact (and unique) solution to the above equation, detailed in Ref.~\cite{JS99} and
obtained with the help of dimensional regularization, reads
\bes
\begin{align}
\phi &= \frac{4G}{c^2} \, \sum_{a} \frac{\alpha_a}{r_{a}}\,,\,\,\,\,\mbox {where}\\
\alpha_a &= \frac{m_a-m_b}{2}         
+ \frac{c^2r_{ab}}{G} \biggl \{ 
\biggl [
1+\frac{m_a+m_b}{c^2r_{ab}/G} +
\no
\left(\frac{m_a-m_b}{2c^2r_{ab}/G}\right)^2 
\biggr ]^{1/2}
 - 1 
\biggr \}\,
\label{Eq_Sk_9b}
\end{align}
\ens
and $r_a$ denotes the Euclidean distance between $x^i$ and $x^i_a$
and $r_{ab}$ the Euclidean distance between $x^i_a$
and $x^i_b$.

   The Brill-Lindquist potential function,
computed in Refs.~\cite{JS99},
reads 
\begin{align}
\label{Eq_Sk_10}
H_{\rm BL}= (\alpha_1 + \alpha_2)~c^2 =(m_1 + m_2)~c^2 - G~ \frac{\alpha_1\alpha_2}{r_{12}}\,,
\end{align}
and as expected it describes the total (initial) energy between two uncharged Brill-Lindquist
black holes. 


The Skeleton approach to General Relativity for dealing with BBH space-times, detailed Ref.~\cite{FJS},
requires both the conformal
flat condition/truncation for the spatial 3-metric and 
hereof together with Eq.~(\ref{Eq_Sk_3b}) 
the maximal slicing condition. These 
prescriptions, valid for all times,
are  given  by  
\begin{subequations}
\begin{align}
\gamma_{ij}& = (1+\frac{1}{8}\phi)^4 \delta_{ij}\,,
\\
\pi^{ij}\gamma_{ij} & = 2 \sqrt{\gamma} \gamma^{ij} K_{ij} = 0\,.
\end{align}
\end{subequations}
In above equations,  $K_{ij}$ is the extrinsic
curvature of $\Sigma(t)$ and 
is related to $\pi^{ij}$ by 
$\pi^{ij} = - \gamma^{1/2}
(\gamma^{il} \gamma^{jm} - \gamma^{ij} \gamma^{lm}) K_{lm}$.

Under the above conditions, the momentum constraint equations for the Einstein theory  
become
\begin{align}
\pi^{j}_{i,\,j} = - \frac{8 \pi G}{c^3} \sum_a p_{ai}\delta_a\,.
\end{align}

The solution of this equation is constructed under the condition that
$\pi^{j}_{i}$ is of purely longitudinal form
\begin{equation}
\label{pij_Sk_def}
\pi^{j}_{i} = \partial_i V_j +  \partial_j V_i - \frac{2}{3}
\delta_{ij} \partial_l V_l\,,
\end{equation}
where $V_i$ is a vector in the flat 3-space. 
The difference to the way $\tilde \pi^{ij}$ is defined, via Eq.~(\ref{piV_def}),
should be noted. In the Skeleton approach, $ \pi^{i}_{j}$ consists only of its
longitudinal part and is defined in terms of $V_i$ via Eq.~(\ref{pij_Sk_def}).
The Hamiltonian constraint in the Skeleton scheme becomes
\begin{align}
\label{Lap_phi}
\Delta\phi &= - \frac{ \pi^j_i \pi^i_j }{(1+\frac{1}{8}\phi)^{7} }
-\frac{16\pi G}{c^2}\sum_a \frac{ m_a\delta_a}{ (1+\frac{1}{8}\phi)}\,
\biggl ( 1+
\nonumber
\\
& \quad
\frac{p_a^2}{(1+\frac{1}{8}\phi)^4m_a^2c^2}
\biggr )^{1/2}\,.
\end{align}
Further, one 
implements a truncation to the numerator of the first term, namely $ \pi^j_i \pi^i_j$,
in the following manner
\begin{align}
\label{piS_Sk}
\pi^j_i \pi^i_j &\equiv -2 V_j\partial_i\pi^i_j + \partial_i(2V_j\pi^i_j) \,
\no
 \rightarrow -2 V_j\partial_i\pi^i_j=\frac{16\pi
G}{c^3}\sum_a p_{aj}V_j\delta_a\,.
\end{align}
This is the other crucial truncation in addition to the conformal flat
truncation required in the Skeleton approach. 
This truncation introduces errors of the same order as the conformal flat
truncation. Therefore, in a post-Newtonian setting, the errors are of the  2PN
order.
With the aid of the ansatz
\begin{align}
\phi = \frac{4G}{c^2}\sum_a \frac{\alpha_a}{r_a}\,,
\end{align}
which is inspired by the delta-function sources in Eq.~(\ref{Lap_phi}), after
the insertion of Eq.~(\ref{piS_Sk}), the resulting energy constraint equation
can be solved in the sense of
dimensional  regularization, detailed in Ref.~\cite{FJS}.
This results in two coupled 
algebraic equations given by 
\begin{align}
\label{eq:alpha_A}
\alpha_a &= \frac{m_a}{1+ \lambda \,\frac{\alpha_b}{r_{ab}}}
\nonumber
\\
& \quad
\times
\left[1 + \frac{p_a^2/(m_a^2 c^2)}{\left(1+ \lambda 
\frac{\alpha_b}{r_{ab}} \right)^{4}} \right]^\frac{1}{2}
\nonumber
\\
& \quad
+ \frac{p_{ai} V_{ai}/c}{\left(1+ \lambda \,
\frac{\alpha_b}{r_{ab}} \right)^{7}}\,,
\end{align}
where  $a=1,2,\,\,b \neq a$ and $\lambda $ stands for 
$ G/(2\,c^2)$.
The $\alpha_a$-parameters generalize the rest energies of
interacting black holes at rest, given in Eq.~(\ref{Eq_Sk_9b}), to orbiting
black holes.  

   With these inputs, the Skeleton Hamiltonian for BBH  becomes
\begin{align}
H_{\rm Sk} \equiv  - \frac{c^4}{16\pi G}\int d^3x\,\Delta\phi = c^2 \sum_a \alpha_a\,.
\end{align}
The Hamilton equations of motion, as usual, read  
\begin{align}
\dot{ \vek x}_a = \frac{\partial H}{\partial \vek p_{a}}\,, \qquad \dot{\vek p}_{a} =
- \frac{\partial H}{\partial \vek x_a}\,.
\end{align}
For a binary system in the center-of-mass frame, we have 
\begin{subequations}
\begin{align}
\vek {p_{1}} &= - \vek{p_{2}} \equiv \vek{p}\,,\\
\vek{r} & \equiv  \vek{x_1}- \vek{x_2}\,,\\
r^2 &= \left ( \vek{x_1}- \vek{x_2} \right) \cdot \left ( \vek{x_1}- \vek{x_2} \right )\,.
\end{align}
\end{subequations}

 Further, we will employ the following convenient dimensionless quantities 
\begin{subequations}
\label{Eq_21_Sk}
\begin{align}
\hat t &= \frac{t\, c^3}{G\,m}\,,\,\,\, \hat r = \frac{r\, c^2}{G\,m}\,,\,\,\, 
\vek{\hat p} = \frac{\vek{p}}{\mu\,c}\,,
\,\,\, {\cal \hat H}_{\rm Sk} = \frac{H_{\rm Sk}}{\mu\,c^2}\,,\\
\hat j & = \frac{J\,c}{G\,m\, \mu}\,,
\,\,\, \hat p_r = \frac{{p_r}}{\mu\,c}\,,
\,\,\, \vek{\hat p}^2 = \hat p_r^2 + \hat j^2/\hat r^2\,,
\end{align}
\end{subequations}
where $\vek{J} = \vek{r} \times \vek{p}$ and $p_r = \vek{p} \cdot \vek{r}/r$.

  For the purpose of present investigation and using Eqs.~(5.1),(5.2) and (5.4) in Ref.~\cite{FJS},
we write the Skeleton Hamiltonian
${\cal \hat H}_{\rm Sk}$ in the following way
\bes
\label{Eq22_Sk}
\begin{align}
{\cal \hat H}_{\rm Sk} &= {2\, \hat r} \biggl ( \psi_1 + \psi_2 - 2 \biggr )
\,,{\rm where}\\
\psi_{1} &=
1+\frac{ \chi_{-} }{4\, \hat r\, \psi_2} \,
\biggl (
1+ \frac{ 4\,{\eta}^{2}
\left( {{\hat p_r}}^{2}+{\frac {{\hat j}^{2}}{{\hat r}^{2}}} \right) }{ \chi_{-}^2\, \psi_2^4}
 \biggr )^{1/2}
\nonumber
\\
& \quad
- \frac{
\left( 8\,{{ \hat p_r}}^{2}+7\,{\frac {{\hat j}^{2}}{{\hat r}^{2}}} \right) {\eta}^{2}}{8\,
{\hat r}^{2}{\psi_{{2}}}^{7} }
\,,\\
\psi_{2} &=
1+\frac{ \chi_{+} }{4\, \hat r\, \psi_1} \,
\biggl (
1+ \frac{ 4\,{\eta}^{2}
\left( {{\hat p_r}}^{2}+{\frac {{\hat j}^{2}}{{\hat r}^{2}}} \right) }{ \chi_{+}^2\, \psi_1^4}
\biggr )^{1/2}
\nonumber
\\
& \quad
- \frac{
\left( 8\,{{ \hat p_r}}^{2}+7\,{\frac {{\hat j}^{2}}{{\hat r}^{2}}} \right) {\eta}^{2}}{8\,
{\hat r}^{2}{\psi_{{1}}}^{7} }
\,,
\end{align}
\ens
and the symbols $ \chi_{-/+} $ are defined to be
$ \chi_{-} = \left( 1-\sqrt {1-4\,\eta} \right) $ and
$\chi_{+} = \left( 1+\sqrt {1-4\,\eta} \right)$.

  The conservative Skeleton Hamiltonian has the following features and properties.
The Skeleton Hamiltonian is exact in the test-body limit, where it
describes the motion of a test particle in the Schwarzschild spacetime.
It is also identical to the 1PN accurate Hamiltonian 
describing 
the compact binary
dynamics in General Relativity. Further, as explained earlier,
when the point particles are at rest, the Brill-Lindquist initial value
solution is reproduced. 
It is remarkable that   
the Skeleton Hamiltonian
allows a post-Newtonian expansion in powers of $1/c^2$ to arbitrary
orders.
This is remarkable because of the observation that the 3PN approximate binary black hole 
Hamiltonian does not exist, if one only imposes the conformal flat condition
without invoking Eq.~(\ref{piS_Sk}) \cite{FJS}. 
Therefore, the Skeleton Hamiltonian describes the
evolution of punctures under both conformal flat conditions for the
3-metric and analyticity conditions for the Hamiltonian 
via PN expansions (and, of course,
gravitational radiation reaction is  not incorporated). 
The Skeleton dynamics is not isolated from the metric field rather it is
embedded in the field [recall that the Skeleton Hamiltonian 
originates from $\gamma_{ij}$].
Furthermore, with the aid of similar truncations, it is also possible to 
calculate $\alpha$ and $\beta^i$ \cite{FJS} and, 
as expected, the demonstration of the equality of ADM and Komar masses 
for BBH in circular orbits in the Skelton approach requires 
explicit expressions for $\alpha$ and $\beta^i$.
These arguments imply that 
in the Skelton scheme for BBH, a global spacetime
picture exists.

\subsection{Tackling the effects of radiation reaction}

 This subsection details how we incorporate radiation reaction effects into the
above described  conservative binary black hole dynamics. Our prescription is heavily influenced
by the way the reactive 2.5PN Hamiltonian 
affects the conservative 2PN accurate compact binary dynamics.
Therefore, let us first take a closer look at the way reactive contributions to ${\cal H}$
change Hamiltonian equations of motion in the post-Newtonian approximation with the help
of 2.5PN accurate ${\cal H}$, available in Refs.~\cite{GS85,DS88}.
It is convenient (and usual) to split the fully 2.5PN accurate ${ H}$ as
\begin{align}
{ H}_{\rm 2.5PN} = { H}^{\rm c}_{\rm 2PN} + { H}^{\rm r}_{\rm 2.5PN}\,,
\end{align}
where ${ H}^{\rm c}_{\rm 2PN}$ defines the 2PN accurate conservative 
orbital dynamics \cite{DS88}. 
The explicitly time dependent reactive Hamiltonian, whose first contribution 
appears at 2.5PN order, is denoted by ${ H}^{\rm r}_{\rm 2.5PN}$.
In terms of the un-scaled position and momentum vector components in the 
center-of-mass frame, and following
Ref.~\cite{GS85}, it reads
\begin{align}
\label{GS85_Hr}
 { H}^{\rm r}_{\rm 2.5PN} =
\frac{2G}{5c^5}\frac{d^3Q_{ij}}{dt^3}\left(\frac{p_ip_j}{\mu}-GM
\mu\frac{r^ir^j}{r^3}\right)
\,,
\end{align}
where $Q_{ij} = \mu\,( r_ir_j - r^2 \, \delta_{ij}/3) $ 
stands for the Newtonian accurate mass-quadrupole of a compact binary.
However, for the purpose of our investigation, it is convenient to consider 
the following physically equivalent reactive Hamiltonian at the 2.5PN order: 
\begin{align}
\label{GS_25r_N}
 {H}^{\rm r}_{\rm 2.5PN}
=  \frac{G\mu}{5c^5}\frac{d^5Q_{ij}}{dt^5}r^ir^j
\,.
\end{align}
The above two reactive Hamiltonians differ by a total time derivative and thus
belong to two different coordinate conditions  
[it is only for  simplicity that dynamical variables are still denoted by same symbols
in Eqs.~(\ref{GS85_Hr}) and (\ref{GS_25r_N})].
The reactive  Hamiltonian given by Eq.~(\ref{GS85_Hr}) is valid
under the ADM coordinate conditions, whereas Eq.~(\ref{GS_25r_N}) is for its 
counterpart in the Burke-Thorne gauge \cite{MTW}(notice that in the Burke-Thorne gauge, the 
reactive part of $h^{\rm TT}_{ ij}$ is zero as explained in Ref.~\cite{GS83}).
However, a change of coordinates at the reactive  order, in comparison with  
the conservative part, does not influence any observable
quantities.     

  With the help of Eq.~(\ref{GS_25r_N}), we define 
${\cal \hat H}^{\rm r}_{\rm 2.5PN} = {H}^{\rm r}_{\rm 2.5PN}/\mu\,c^2 $, 
and obtain, rather symbolically for the 2PN conservative part,
the following fully 2.5PN accurate Hamiltonian equations of motion 
in polar coordinates and in dimensionless  
variables  as
\begin{subequations}
\label{H25_341}
\begin{align}
\frac{d\hat r}{d \hat t} &= \frac{\partial {\cal  \hat H}^{\rm c}_{\rm 2PN} } {\partial \hat p_r }\,,\\
\frac{d \phi}{d \hat t} 
&= \frac{\partial {\cal \hat H}^{\rm c}_{\rm 2PN}  } {\partial \hat j }
\equiv \hat \omega  
\label{H25_341b} \\
\frac{d \hat p_r}{d \hat t}  &= -\frac{\partial {\cal \hat H}^{\rm c}_{\rm 2PN}  } {\partial \hat r } + f_{\hat p_r}\,,
\label{H25_341c} \\
\frac{d \hat j}{d \hat t}  &= -\frac{\partial {\cal \hat H}^{\rm r}_{\rm 2.5PN}  } {\partial \phi  } \equiv f_{\hat j}\,,
\end{align}
\end{subequations}
where ${\cal \hat H}^{\rm c}_{\rm 2PN} = {H}^{\rm c}_{\rm 2PN}/\mu\,c^2 $ and
the explicit expressions for $f_{\hat p_r}$ and $f_{\hat j}$ are given by
\begin{subequations}
\label{fprfj}
\begin{align}
f_{\hat p_r} &=
\frac {16 \, \eta \, \hat p_r }{15\, \hat r^5}\, \biggl \{  4\, \hat r
+3\,{{\hat p_r}}^{2}{\hat r}^{2}+18\,{\hat j}^{2} \biggr \} 
\,,\\
f_{\hat j} &=
-\frac{8\,\eta\,\hat j}{5\,\hat r^5}\, \biggl \{  -9\,{{\hat p_r}}^{2}{\hat r}^{2}+6\,{\hat j}^{2}
-2\,\hat r \biggr \}  
\,.
\end{align}
\end{subequations}
 As mentioned earlier, in this investigation we focus on black hole binaries 
inspiralling along quasi-circular orbits and we 
prescribe the quasi-circularity by imposing $\hat p_r \equiv 0$ on the
reactive contributions to Hamiltonian equations of motion. 
This implies that Eqs.~(\ref{fprfj}) becomes
\begin{subequations}
\label{fprfj_c}
\begin{align}
f_{\hat p_r} & \equiv  0
\,,\\
f_{\hat j} &= - \frac{ 32}{5} \, \frac{\eta }{ \hat r^{7/2}} 
\,.
\end{align}
\end{subequations}
A close look at the above equation for $f_{\hat j}$ 
that defines $d\hat j/dt$ reveals
that it is identical to
the dominant contribution to  the far-zone angular momentum flux associated with comparable mass 
binaries inspiralling along quasi-circular orbits.
This observation 
helped us to provide the following prescription to incorporate radiation reaction 
on the conservative Skeleton Hamiltonian, given by Eqs.~(\ref{Eq22_Sk}). 

 We recall that the Skeleton Hamiltonian providing the conservative part of the binary black hole
dynamics is PN independent and therefore can be 
employed to treat close BBH configurations, provided we accept the validity of
underlying assumptions.
Therefore,  in order to introduce gravitational radiation reaction 
effects,  
we invoke the most PN accurate expression for 
far-zone angular momentum flux associated with comparable mass compact
binaries inspiralling along quasi-circular orbits to provide an evolution equation for $\hat j$. 
This is how we incorporate radiation reaction into Eqs.~(\ref{Eq22_Sk}) and the dynamics it defines.
The fully 3.5PN accurate expression for far-zone angular momentum flux for compact binaries
in quasi-circular orbits, denoted by $L_j$ and extractable from Ref.~\cite{3PN_res}, reads
\begin{align}
\label{djdt_35PN}
L_j &= 
{\frac {32}{5}}\,\eta\,{v}^{7} 
\biggl \{
 1
- \biggl [ {\frac {1247}{336}}+ {\frac {35}{12}}\,\eta  \biggr ] {v}^{2}
+4\,\pi\,{v}^{3}
\nonumber
\\
& \quad
+  \biggl [ -{ \frac {44711}{9072}}+{\frac {9271}{504}}\,\eta
+{\frac {65}{18}}\,{\eta
}^{2}  \biggr ] {v}^{4}
- \biggl [ {\frac {8191}{672}}
\nonumber
\\
& \quad
+{\frac {583}{24}}\,
\eta  \biggr ] \pi\,{v}^{5}
+ 
\biggl [
 {\frac {6643739519}{69854400}}+ \frac{16}{3}
\,{\pi }^{2}
\nonumber
\\
& \quad
-{\frac {1712}{105}}\,\gamma+ \left( -{\frac {134543}{7776
}}+{\frac {41}{48}}\,{\pi }^{2} \right) \eta
-{\frac {94403}{3024}}\,{ \eta}^{2}
\nonumber
\\
& \quad
-{\frac {775}{324}}\,{\eta}^{3}-{\frac {1712}{105}}\,\ln 
 \left( 4\,v \right)  
\biggr ]
 {v}^{6}
+ \biggl [
-{\frac {16285}{ 504}}
\nonumber
\\
& \quad
+{\frac {214745}{1728}}\,\eta+{\frac {193385}{3024}}\,{\eta}^{2}
\biggr ]\, \pi\,
{v}^{7} 
\biggr \}\,,
\end{align}
where $ v \equiv \hat \omega^{1/3} $ and  
$\hat \omega = d \phi/d\hat t $ is the conservative dimensionless orbital frequency.
The above equation is derived by noting that for circular orbits, the far-zone angular momentum 
and energy fluxes are related by $ L_E = \omega L_j$.
 
 Therefore, the Skeleton dynamics representing dynamics of comparable-mass non-spinning 
coalescing black hole binaries, in the scaled polar coordinates, can be summarized as 
\begin{subequations}
\label{Sk_341}
\begin{align}
\frac{d\hat r}{d \hat t} &= \frac{\partial \cal  \hat H_{\rm Sk}} {\partial \hat p_r }\,,\\
\frac{d \phi}{d \hat t} 
 &= \frac{\partial \cal \hat H_{\rm Sk}} {\partial \hat j }
\equiv \hat \omega 
\,,
\label{Sk_341b} \\
\frac{d \hat p_r}{d \hat t}  &= -\frac{\partial \cal  \hat H_{\rm Sk}} {\partial \hat r }\,,
\label{Sk_341c} 
\\
\frac{d \hat j}{d \hat t}  &= -L_{j}\,,
\end{align}
\end{subequations}
where $L_j$, the 3.5PN accurate far-zone angular momentum flux, is given by Eq.~(\ref{djdt_35PN}). 
It is important to note that, in our prescription,
$\hat \omega$ appearing in the PN accurate  expression for 
$L_j$ via $v \equiv \hat \omega^{1/3} $ is provided by the 
Skeleton dynamics [in other words,  by Eq.~(\ref{Sk_341b})].
The computational details required to compute right hand sides of Eqs.~(\ref{Sk_341})
are provided in the Appendix~\ref{Sk:Appendix:A}

  We have already argued that the Skeleton Hamiltonian, given by Eq.~(\ref{Eq22_Sk}),
does provide a certain conservative dynamics for two punctures orbiting each other.
Therefore, Eqs.~(\ref{Sk_341}) represent a prescription
to implement semi-analytically puncture evolution to model 
merging binary black holes with reasonable precision.
It should be noted that 
the use of Eq.~(\ref{djdt_35PN}) implies that the adiabatic inspiral occurs along
quasi-circular orbits.


  We observe that the arguments that allowed us to employ PN accurate far-zone angular momentum flux
to incorporate the effects of radiation reaction into our conservative (and PN independent)
Hamiltonian is fairly elegant compared to somewhat heuristic arguments employed by 
the advocates of the EOB scheme (see Sec.~III in Ref.~\cite{BD00}).
Further, the EOB approach usually requires re-summed Pad\'e estimates for $d \hat j/d \hat t$ 
that also require 
PN accurate far-zone angular momentum flux, to include effects of radiation reaction on 
the conservative EOB Hamiltonian. We have experimented with various 
Pad\'e estimates for $d \hat j/d \hat t$ in our Skeletonian approach. However, the final value of 
the Skeletonian $\hat j$, when $\hat \omega$ reaches its maximum value,
turned out to be rather insensitive to the choice of
Pad\'e or Taylor approximant for $d \hat j/d \hat t$.
Therefore, in this paper, we 
employed only the usual 3.5PN (Taylor) accurate  expression for 
far-zone angular momentum flux, in terms of $v \equiv \hat \omega^{1/3}$, given by Eq.~(\ref{djdt_35PN}).

 In the next section, we describe how we numerically implement the reactive Skeleton 
dynamics and explore its various facets. 

\section{Numerical Exploration of reactive Skeleton dynamics } 
\label{Skdyn_results}

 Let us first detail how we provide initial conditions required to define 
${\cal \hat H}_{\rm Sk}$ and to solve Eqs.~(\ref{Sk_341}).

\subsection{Initial conditions required to define reactive Skeleton dynamics}

It is obvious that the equations describing the late stages of 
binary black hole dynamics, given by Eqs.~(\ref{Sk_341}),
require us to specify ${\cal \hat H}_{\rm Sk}$ defined in terms of 
$\psi_1(\psi_2,\hat r,\hat p_r,\hat j)$ and $\psi_2(\psi_1,\hat r,\hat p_r,\hat j)$. 
This implies that we need to find ways to
specify values for $\hat r,\phi, \hat p_r, \hat j, \psi_1$ and $\psi_2$
at an initial instant, say $t=0$.
The initial conditions for $\hat r$ and $\phi$ are arbitrary and specify the initial radial 
separation and its associated angular position. 

  The 2PN accurate expression for $\hat p_r$ in 
terms of $\hat r$ in the ADM coordinates, displayed below, is 
our initial condition for $\hat p_r$
\begin{align}
\label{pr_ini}
\hat p_r = &
-{\frac {64}{5}}\,{\frac {\eta\,}{{\hat r}^{3}}}
+ \left( {\frac {460} {21}}\,{\frac {\eta}{{\hat r}^{4}}}
+16\,{\frac {{\eta}^{2}}{{\hat r}^{4}}} \right) 
-{\frac { 256}{5}}\,{\frac {\eta\,\pi}{{\hat r}^{9/2}}}
\nonumber
\\
& \quad
+ \left( -{\frac {1568}{45}}\,{\frac {{\eta}^{2}}{{\hat r}^{5}}}-{\frac {
158224}{2835}}\,{\frac {\eta}{{\hat r}^{5}}} \right) 
\,.
\end{align}
The derivation of Eq.~(\ref{pr_ini}) requires following inputs and is done in three steps.
First, we compute $d\hat r/d \hat t$ associated with the quasi-circular inspiral to 2PN order with the help
of Ref.~\cite{3PN_res}. Employing the 2PN accurate relation connecting $\hat j$ and $\hat r$, relevant for 
PN accurate circular orbits, we iterate the 2PN accurate Hamiltonian Equation for 
symbolically
$d\hat r/d \hat t$, written in terms of $\hat r,\hat p_r$ and $\hat j$, and compute $\hat p_r$ 
as a PN accurate polynomial
in $d \hat r/d\hat t$. In the final step, we invoke $d \hat r/d \hat t$, accurate to 2PN radiation reaction order 
(and was computed in the first step) and obtain PN accurate expression for $\hat p_r$
in terms of $\hat r$, given by Eq.~(\ref{pr_ini}).

 The initial value for $\hat j$ is computed 
by solving $d \hat p_r/d \hat t=0$, where $d \hat p_r/d \hat t$ is given by Eq.~(\ref{Sk_341c}).
It should be obvious that we need to specify $r, p_r, \psi_1$ and $\psi_2$ to obtain numerically
the value for $ \hat j$ by imposing the right hand side of  Eq.~(\ref{Sk_341c}) to zero.
We provide initial guesses for $\psi_1$ and $\psi_2$ by invoking the ingredients
used to construct Brill-Lindquist potential function, associated with
the Brill-Lindquist initial value solution for two black holes at rest \cite{JS99}.
This leads to the following expressions for $\psi_1$ and $\psi_2$ in terms of
$\hat r$ and $\eta$:
\begin{subequations}
\label{psi_BL}
\begin{align}
\psi_{1}^{\rm BL} =&
\frac{1}{2}-{\frac {\sqrt {1-4\,\eta}}{4\,\hat r}}+\frac{1}{2}\,
\sqrt { 1+\frac{1}{\hat r}+\frac{1}{4\,\hat r^{2}}-{\frac {\eta}{{\hat r}^{2}}} }
\,,\\
\psi_{2}^{\rm BL} =&
\frac{1}{2}+{\frac {\sqrt {1-4\,\eta}}{4\,\hat r}}+\frac{1}{2}\,
\sqrt { 1+\frac{1}{\hat r}+\frac{1}{4\,\hat r^{2}}-{\frac {\eta}{{\hat r}^{2}}} }
\,.
\end{align}
\end{subequations}
We would like to emphasize that it is only during the evaluation of an initial guess for $\hat j$ that we invoke 
Eqs.~(\ref{psi_BL}).
Further, while performing various checks, 
we realized that the initial value of $\hat j$ is rather insensitive to the choice of $\hat p_r$.
In other words, initial value for $\hat j$ does not change much even when we use $\hat p_r=0$ 
on the right hand side of
Eq.~~(\ref{Sk_341c}).

  We are now in a position to obtain initial values for $\psi_1$ and $\psi_2$ that are required to 
solve Eqs.~(\ref{Sk_341}). The initial values for  $\psi_1$ and $\psi_2$, associated with 
the above mentioned $\hat r,\hat p_r$ and $\hat j$ values are numerically evaluated using the coupled
algebraic equations for $\psi_1$ and $\psi_2$, given in Eqs.~(\ref{Eq22_Sk}).

  From here onwards, we drop {\it hat symbol} appearing on 
our dimensionless quantities like ${\cal \hat H_{\rm Sk} }, \hat r, \hat p_r, \hat j, \hat \omega, \hat t $ 
for the ease of presentation.

\subsection{Various aspects of the reactive Skeleton binary black hole dynamics  }

 Before we probe numerically the various aspects of the SAPE  
to model merging binary black holes, having $\eta = 1/4$, let us state that 
the way of solving Eqs.~(\ref{Sk_341}) is quite different 
from the way one solves Hamiltonian equations when the Hamiltonian is explicitly given in terms 
of $r,p_r$ and $j$. 
While dealing with the reactive 2.5PN accurate Hamiltonian system, given by Eqs.~(\ref{H25_341}),
or the EOB scheme,
one needs to invoke \emph {only once} a numerical scheme,
like the $4^{\rm th}$ order Runge-Kutta, to solve the associated couple 
differential equations, after specifying certain initial conditions for $r,p_r$ and $j$.
However, a similar approach is not applicable for the reactive Skeleton dynamics, given by
Eqs.~(\ref{Sk_341}) .
This is because of the fact that the Skeleton Hamiltonian, given by Eq.~(\ref{Eq22_Sk}), implicitly
depends on $r,p_r$ and $j$ and its explicit dependence on  $\psi_1$ and $\psi_2$ 
is interconnected. 
Therefore, the numerical solution to Eqs.~(\ref{Sk_341}) is determined in the following way.
We have already detailed how to prescribe initial values for 
$ r, \phi, p_r, j, \psi_1$ and $\psi_2$ in the previous subsection.
With the aid of these initial guesses, we invoke the $4^{\rm th}$ order Runge-Kutta to solve 
Eqs.~(\ref{Sk_341}) and 
obtain values for
$r,p_r,\phi$ and $j$ at the next time step $\delta t$. 
These values are employed  to evaluate the values of $\psi_1$ and $\psi_2$ at the instant $\delta t$,
with the help of 
the coupled
algebraic equations for $\psi_1$ and $\psi_2$, given in Eqs.~(\ref{Eq22_Sk}).
To obtain the values of the dynamical variables at the next instant, $2 \delta t$, we use as initial guesses,
the values of $r, p_r, \phi, j,  
\psi_1$ and $\psi_2$ at the earlier step, and repeat the 
above detailed procedure all over again till we reach the terminal 
point of our evolution.
This is how we obtain numerical solution to the reactive Skeleton dynamics for 
binary black hole evolution, given by Eqs.~(\ref{Sk_341}).
In other words, we need to call the $4^{\rm th}$ order Runge-Kutta
at every time step $t$ to obtain the values of 
$r, p_r, \phi, j,
\psi_1$ and $\psi_2$ at $ t + \delta t$. However, while solving the 2.5PN accurate Hamiltonian equations,
Eqs.~(\ref{H25_341}) or the EOB differential equations, one usually invokes a numerical scheme like 
the $4^{\rm th}$ order Runge-Kutta only once.

 We are now in a position to extract (numerically) various aspects of 
the reactive Skeleton dynamics for
binary black hole evolution. 
In this paper, we 
terminate the SAPE when $\omega(t)$ reaches its maximum value, $\omega_{\rm mx}$  and 
let the initial value for $r$ to be $10$. 
Plots depicting temporal evolutions of various dynamical variables, appearing in Eqs.~(\ref{Sk_341}),
for a binary black hole having $\eta=0.25$ is presented in Fig.~\ref{fig:xyEjpr}.
The parametric plot of $x = r\, \cos \phi$ versus $y= r\, \sin \phi$ clearly indicates 
that the dynamical evolution, predicted by Eqs.~(\ref{Sk_341}), occurs along inspiralling
quasi-circular orbits and it is rather difficult to pin-point the position of the last stable orbit,
which can clearly be defined using the conservative Skeleton Hamiltonian as done 
in Ref.~\cite{FJS}. Further, we observe that both the binding energy, given by the Skeleton 
Hamiltonian, and $j$, the orbital angular momentum vary smoothly till we terminate the
reactive Skeleton evolution at $\omega = \omega_{\rm mx} \sim 0.0896 $.
The temporal behavior of $p_r$ clearly indicates that even in the neighborhood of
$\omega_{\rm mx}$, the orbital motion is not substantially different from a quasi-circular inspiral.
The region where $p_r$ increases sharply may considered to be the plunge phase. 
In Fig.~\ref{fig:omgtomgrrtdrSt}, we explore $r(t), \omega(t) $ and $\dot r^2 (t) $ under 
reactive Skeleton dynamics. Plots of $\omega$ versus $t$ and  $\omega$ versus $r$ indicate
faster evolution for $\omega$ after the late inspiral stage. The rapid decrease in $r$
during the plunge phase is also evident in the  $r$ versus $t$ plot.
Interestingly, $\dot r^2 = (dr/dt)^2 $ remains small even during the plunge phase.
However, it changes from $\sim 10^{-3}$ to $\sim 10^{-2}$ when $t$ changes from $\sim 588$ to $\sim 627$,
which clearly indicate deviations from the adiabatic evolution along circular orbits of
decreasing radii, expected during the quasi-circular inspiral.
Another interesting aspect of the SAPE is that the reactive evolution becomes 
slightly eccentric if we choose $p_r \equiv 0$ at the initial instant $t=0$ and 
this is also displayed in Fig.~\ref{fig:omgtomgrrtdrSt}.
We would like to point out that residual eccentricities appearing in the 
puncture evolution in full General Relativity were suppressed in a similar manner in Ref.~\cite{HHBGS}.

 The nature of the BBH evolution during the plunge phase, predicted by Eqs.~(\ref{Sk_341})
is probed in Fig.~\ref{fig:tVS_KC_QC}.
First, we plot the $\omega^2 \,r^3 $, the `Kepler combination', against $r$.
This plot is motivated by the observation in Ref.~\cite{DG06} that for the reactive EOB evolution
$\omega^2 \,r^3 $, which remains close to unity during the quasi-circular inspiral,
decreases rapidly during the plunge.
We note that, similar to what is observed in the Numerical Relativity
based evolution of punctures, the 
`Kepler combination' varies even during the inspiral. 
Therefore, it is reasonable to state that the constancy of `Kepler combination', observed in
Ref.~\cite{DG06},
is purely a consequence of the fact that the EOB scheme employs Schwarzschild-like coordinates.
In Fig.~\ref{fig:tVS_KC_QC}, we also plot $v_{\omega}^2 \equiv \omega^{2/3}$ versus $t$
and $v_{\rm G}^2 \equiv r^2\,\omega^2 + \dot r^2 $ versus $t$ and 
observe that $v_\omega $ remains larger than $v_{\rm G}$ throughout the Skeleton 
inspired inspiral. The above two plots indicate that in reactive Skeleton dynamics,
it is not very attractive to employ $v= r\, \omega$ or $ v= v_{\rm G}$
in the expression for the far-zone angular momentum flux while incorporating
radiation reaction on the conservative Skeleton dynamics.
A plot of the `quasi-circularity' condition
$d{ E} = \bar \omega dj $, valid for binary inspirals along a sequence of circular orbits
is also available in Fig.~\ref{fig:tVS_KC_QC}.
The quantities $dE$ and $dj$ represent differences in the values of $\cal H_{\rm Sk}$ and 
$j$ at $t$ and $t+ \delta t$, while $\bar \omega$ is the average value of $\omega$, given by
Eq.~(\ref{Sk_341b}), at $t$ and $t+ \delta t$.
We observe that the degree with which the above relation is violated keeps getting 
bigger as we approach $\omega_{\rm mx}$. However, the difference between 
$dE$ and $\bar \omega\, dj$ is small even in the neighborhood of $\omega_{\rm mx}$ indicating 
that the plunge motion is fairly close to `quasi-circular'.
Therefore, in our opinion, the plots in Fig.~\ref{fig:tVS_KC_QC} suggest that it is not very
unrealistic to employ $v_{\omega}$ to estimate the effects of radiation reaction in
our SAPE.


   Let us perform a first-order comparison between the GW phase evolutions
based on SAPE and Numerical Relativity (NR) for an equal mass binary during its late inspiral
without using any data from NR simulations.
This is possible due to the observation, reported in Ref.~\cite{CC07}, that 
GW phase evolution associated with certain PN approximant, namely TaylorT4 at 3.5PN order,
closely agrees with 
equal mass binary black hole NR simulations that last around 15 orbits prior 
to the merger[the accumulated phase difference is less than 0.05 radians over the 30-cycle waveform]. 
The orbital phase evolution for equal mass binary black holes under TaylorT4 
approximant at 3.5PN order is obtained by numerically 
integrating the following two differential equations \cite{CC07}:
\begin{subequations}
\label{EqP4}
\begin{align}
\label{EqP4a}
\frac{d \phi (t)}{dt} &\equiv  x^{3/2}\, ,\\
\frac{d\,x(t)}{dt} &=
\frac{16}{5\,}\, x^5\, \biggl \{
1
-{\frac {487}{168}}\,x
+4\,\pi\,{x}^{3/2}
\noQ
&
+{ \frac {274229}{72576}}\,{x}^{2}
-{\frac {254}{21}}\,\pi\,{x}^{5/2}
\noQ
&
+
\biggl [
{\frac {178384023737}{
3353011200}}-{\frac {1712}{105}}\,\gamma+{\frac {1475}{192}}\,{\pi}^{2
}
\noQ
&
-{\frac {856}{105}}\,\ln \left( 16\,x \right)
\biggr ]
 {x}^{3}
+{\frac {3310}{189}}\,\pi\,{x}^{7/2}
\biggr \}
\,,
\label{EqP4b}
\end{align}
\end{subequations}
where $\gamma$ is the Euler gamma and $x =\omega^{2/3}$.
In what follows, we compare the orbital phase evolution prescribed by SAPE
for $\eta =0.25$ with $\phi(t)$ originating from Eq.~(\ref{EqP4}).

 There are two slightly different ways to compare $\phi_{\rm Sk}(t)$ against
$\phi_{\rm T4}(t)$ after choosing a specific interval for $x$, defined
by some minimum and maximum values, say $x_{\rm min}$ and $x_{\rm max}$, for $x$.
In the first approach, one lines up $\phi_{\rm Sk}(t), \phi_{\rm T4}(t)$
and their time derivatives at $x = x_{\rm max}$ and with the help of 
a shifted $\phi_{\rm T4'}(t)$, given by Eq.~(49) in Ref.~\cite{CC07},
computes $\phi_{\rm Sk}(t)- \phi_{\rm T4'}(t)$ when $x= x_{\rm min}$ [
this is pursued in Refs.~\cite{CC07_3639}].
In the second prescription, one lines up 
$\phi_{\rm Sk}(t), \phi_{\rm T4}(t)$
and their time derivatives at $x = x_{\rm min}$ and computes 
 $\phi_{\rm Sk}(t)- \phi_{\rm T4}(t)$ when $x$ reaches $x_{\rm max}$ and 
this is what we list below and the differences 
between the two prescriptions in our comparisons are always negligible.
In Fig.~\ref{fig:Sk_T4}, we plot $\phi_{\rm Sk}(t)- \phi_{\rm T4}(t)$ in radians
and $x(t)$ under SAPE and T4 approximant at 3.5PN order 
for various $x$ intervals. For the top panel, the orbital frequency range 
is from $\omega = 0.025$ to $\omega= 0,05$ and there are $\sim 8.5$ cycles.
In the middle panel, we vary $\omega$ from $0.025$ to $0.045$ and there are about
$8$ cycles. Finally, for the bottom panel there are about 11.5 cycles and the
$\omega$ interval is from 0.0222 to 0.045. Further,
the fractional differences in $\phi$  are $\sim 7 \%, 5.8\%$ and $\sim 6.1 \% $ 
in the above three cases and it increase sharply as we approach the plunge.

 It should be emphasized that we haven't employed any
arbitrary parameters in our SAPE.
Therefore, it is conceivable that by introducing flexible parameters,
we should be able to obtain a modified reactive Skeletonian dynamics 
that prevents the above observed dephasing.
This is inspired by what is advised in the modified EOB approach \cite{EOB_Golm} 
and it will be reported in a separate publication. 

\section{Rough prescription to obtain BBH coalescence waveforms via SAPE}
\label{Skdyn_h}

 We aim to provide, in this section, a rough estimate for the GW polarizations, having Newtonian
accurate amplitude, associated with the BBH coalescence in our SAPE.
This is achieved by terminating the SAPE when $\omega$, given by Eq.~(\ref{Sk_341b}), reaches its maximum value 
$\omega_{\rm mx}$ an then matching there  
the relevant time derivatives of the radiative multipole moments associated with 
the late plunge phase to the corresponding `ring-down' multipole moments, constructed using
appropriate quasi-normal mode(QNM) contributions.
Our construction of $h_{\times}^{\rm N}(t) $ and $h_{+}^{\rm N}(t)$, {\it i.e.} 
the temporally evolving GW polarizations having Newtonian
accurate amplitude, for the ring-down phase after the termination of the SAPE follows 
very closely what is detailed Sec.~IV of Ref.~\cite{DG06}. 

  The GW polarizations are defined in the following way
\bes
\begin{align}
h_{\times} &=   h^{\rm TT}_{ \vek { \hat \theta} \, \vek {\hat \phi} }
\,,\\
h_{+} &=  \frac{1}{2} \biggl ( 
h^{\rm TT}_{ \vek { \hat \theta} \, \vek {\hat \theta} } 
- h^{\rm TT}_{ \vek { \hat \phi} \, \vek {\hat \phi} }
  \biggr ) \,,
\end{align}
\ens
where $  h^{\rm TT}_{ \vek { \hat \theta} \, \vek {\hat \phi} }$
and $ h^{\rm TT}_{ \vek { \hat \theta} \, \vek {\hat \theta} }$
are the independent  
components of 
$h^{\rm TT}_{ ij}$ representing the radiation field emitted 
by isolated systems in asymptomatically flat space-time. The various angular components of 
$h^{\rm TT}_{ ij}$ are defined using a spherical coordinate system, having orthonormal 
bases $( \vek {\hat r'}, \vek{\hat \theta}, \vek{\hat \phi})$, centered on the center-of-mass of the binary  
and whose  azimuthal axis is aligned with the orbital angular momentum vector ${\vek l}$.
Following Ref.~\cite{KT80}, it is possible to obtain 
$h^{\rm TT}_{ ij}$ in terms of certain time derivatives of the
radiative mass and current moments ${\cal I}^{lm}$ and ${\cal S}^{lm}$, 
irreducibly defined with respect to ${\vek l}$,
where $m = -l, -l+1,..,+l$ and $T^{\rm E2,lm}_{ ij}$ and  $T^{\rm B2,lm}_{ ij}$: the so-called
pure-spin tensor-spherical harmonics of electric and magnetic types.
To compute $h_{\times,+}^{\rm N}$, we only require dominant contributions to 
$h^{\rm TT}_{ ij}$ given by
\begin{align}
h^{\rm TT}_{ ij}|_{\rm N} =
\frac{G}{c^4\, r'} \sum_{m =-2}^{2} \,
^{(2)}{\cal I}^{2m} (t - r'/c) \, T^{\rm E2,2m}_{ij}\,.
\end{align} 
It turns out that at the Newtonian order, ${\cal I}^{20} = {\cal I}^{21} \equiv 0$ 
and contributions arising from $m =-2$ are  
evaluated using the usual relation  ${\cal I}^{2-2} = {\cal I}^{2+2^{*}} $, 
where $*$ stands for  complex conjugation.
Therefore, only the following expressions are required  to compute $h_{\times,+}^{\rm N}$
\bes
\label{EqSk_35}
\begin{align}
^{(2)}{\cal I}^{22} &=
-\frac{8}{5}\,\sqrt{10\, \pi}\, \eta\, m\,c^2\, \omega^{2/3}\, e^{-2\,i\, \phi(t)}
\,,\\
T^{\rm E2,22} &= \left ( \frac{5}{128\, \pi}\right )^{1/2} \,
e^{2\,i\, \varphi}\,
\biggl \{ 
\biggl [( 1 + \cos \vartheta^2 ) 
\biggl ( \vek { \hat \theta} \otimes  \vek {\hat \theta }
\noQ 
&
- \vek { \hat \phi} \otimes  \vek {\hat \phi }
\biggr )
+ 2\, i\, \cos \vartheta 
\biggl ( \vek { \hat \theta} \otimes  \vek {\hat \phi }
+ \vek { \hat \phi} \otimes  \vek {\hat \theta }
\biggr \}
\,.
\end{align}
\ens

 With these inputs, we obtain
\bes
\label{EqSk_36}
\begin{align}
h_{\times}^{\rm N} &= -4\, \eta \, \cos i\, \left ( \frac{G\,m}{c^2\, r'} \right )\, \omega^{2/3}\, \sin 2\,\phi
\,,\\
h_{+}^{\rm N} &= -2\, \eta \, \left ( 1+ \cos^2 i \right )\, 
\left ( \frac{G\,m}{c^2\, r'} \right )\, \omega^{2/3}\, \cos 2\,\phi\,,
\end{align}
\ens
where we identified $\vartheta = i$, the orbital inclination angle and set $\varphi=0$.

 Following Ref.~\cite{DG06}, we aim to match as smooth as possible 
$^{(2)}{\cal I}^{2\pm 2}_{\rm plunge} $, given in Eqs.~(\ref{EqSk_35}), 
at $t= t_{\rm m}$, 
$t_{\rm m}$ being the time when the SAPE $\omega(t)$  
reaches its maximum value, to the corresponding ring-down 
$^{(2)}{\cal I}^{2\pm 2}_{\rm ring}$ consisting of certain sum of decaying QNM modes.
For simplicity, we restrict our attention to 
$^{(2)}{\cal I}^{2\pm 2}_{\rm ring}$ constructed using 
the first two complex conjugated fundamental
QNM modes. For this purpose, we only need to know
the following conjugate pair of complex QNM frequencies,
extractable from Ref.~\citep{KS_LR},
\begin{align}
\sigma^{\pm}_{20} &= 0.08896 \pm 0.37367\, i\,.
\end{align}
Under these restrictions, $^{(2)}{\cal I}^{2\pm 2}$ during the ring-down phase is 
given by
\begin{align}
^{(2)}{\cal I}^{2+2}_{\rm ring} &= 
C^{+}_{0}\, e^{- \sigma^{+}_{20}\, \tau }
+ C^{-}_{0}\,e^{- \sigma^{-}_{20}\, \tau}
\no
\,\,\,\,\, 
( {\rm for  } \,\,\,\tau \equiv t-t_{\rm m} >0  )
\end{align}
The two arbitrary
complex coefficients $C^{+}_{0} ({\cal I}^{22})$ and $C^{-}_{0} ({\cal I}^{22}) $ 
can be chosen so as to ensure  not only that 
$^{(2)}{\cal I}^{22}_{\rm plunge } (t=t_{\rm m})$
agrees with $^{(2)}{{\cal I}^{22}_{\rm ring} }(t=t_{\rm m})
= C^{+}_{0} ( {\cal I}^{22} ) + C^{-}_{0} ( {\cal I}^{22} )$, but also that the
numerically computed
time derivative of $ ^{(2)}{{\cal I}^{22}_{\rm plunge } }(t ) $, {\it i.e.} 
$^{(3)}{{\cal I}^{22}_{\rm plunge } }(t)$
agrees, when $ t= t_{\rm m}$ with
$ ^{(3)}{{\cal I}^{22}_{\rm ring }}
= - \sigma^{+}_{20}\, C^{+}_{0} ( {\cal I}^{22} ) - \sigma^{-}_{20}\, C^{-}_{0} ({\cal I}^{22} ) $.
This yields
\begin{subequations}
\begin{align}
C^{+}_{0} ({\cal I}^{22} ) &= \bigg (  \sigma^{-}_{20} - \sigma^{+}_{20} \biggr )^{-1}\,
 \biggl [
\sigma^{-}_{20} \, 
^{(3)}{ {\cal I}^{22}_{\rm plunge} }(t)
\no
+  
^{(4)}{ {\cal I}^{22}_{\rm plunge } }(t)
\biggr ]_{t=t_{\rm m} }
\\
C^{-}_{0} ( {\cal I}^{22} ) &=
\biggl ( { \sigma^{+}_{20} - \sigma^{-}_{20} } \biggr )^{-1}
\times
\biggl [
\sigma^{+}_{20} \, ^{(3)}{ {\cal I}^{22}_{\rm plunge} }(t)
\no
+  
^{(4)}{ {\cal I}^{22}_{\rm plunge } }(t)
\biggr ]_{t = t_{\rm m} } 
\end{align}
\label{c0_pm_f}
\ens
We would like to point out that while computing 
$^{(3)}{\cal I}^{2+2}_{\rm plunge}(t)$, we include 
the $ d \omega/dt$ contribution,  
with the help of Eqs.~(\ref{dpsidt}).
Using the above detailed prescriptions for $^{(2)}{\cal I}^{2+2}_{\rm ring}$, it is now
fairly straightforward to obtain 
$h_{\times,+}^{\rm N}$ associated with BBH coalescence under SAPE that also includes contributions
from QNM ringing phase.  

  In  Numerical Relativity, it is 
customary to invoke the Weyl scalar $\psi_4$ to represent the gravitational wave content of the 
dynamical space-time. In a suitable null tetrad and in the far-zone, one finds
\begin{align}
\psi_4 = \ddot h_{+} - i\, \ddot h_{\times}\,.
\end{align}  
The $\Re (\psi_4) $ and $\Im (\psi_4)$ having `Newtonian' amplitude, similar to Eqs.~(\ref{EqSk_36}) 
read
\begin{subequations}
\label{EqSk_41}
\begin{align}
\Re (\psi_4) &= -8\,\left (1 +  \cos i^2\, \right )\,
\left ( \frac{G\, m}{c^2\, r'} \right )
\, \eta \, \omega(t)^{8/3}
\, \cos 2\, \phi (t)
\,,\\
 \Im (\psi_4) &=
-16\, \cos i\,
\left ( \frac{G\, m}{c^2\, r'} \right )\,
 \eta \, \omega(t)^{8/3}\,  \sin 2\, \phi (t)\,.
\end{align}
\end{subequations}
   The resulting plots for $h_{\times,+}^{\rm N}(t)$ and the associated $\Re (\psi_4) $ and 
$\Im (\psi_4) $
are displayed in Fig.~\ref{fig:thxppsi4RI}.
The plots clearly demonstrate a smooth matching 
to the QNM ringing
phase in the SAPE.

\section{Discussion and future directions}
\label{Skdyn_CR}

 This paper provides a semi-analytic prescription to model the BBH coalescence where we employ
the punctures, employed by several Numerical Relativity groups, to model the non-spinning black holes.
Our approach requires the conservative Skeleton Hamiltonian, derived in Ref.~\cite{FJS},
representing two orbiting punctures 
in a waveless truncation to the
Einstein field equations.
We include the effects of radiation reaction in a transparent Hamiltonian framework
that fully justifies the physical system we are trying to model.
This is how we provide a prescription to implement our semi-analytic puncture evolution, termed SAPE,
to model the merging binary black-holes.
We also provided GW polarizations, real and imaginary parts of the Weyl scalar, 
having Newtonian amplitude, associated with the entire 
BBH coalescence in our SAPE. This is achieved by  
matching smoothly $^{(2)}{{\cal I}^{2\pm 2}_{\rm plunge } }(t)$, when $\omega = \omega_{\rm mx}$,
to $^{(2)}{ {\cal I}^{2\pm 2}_{\rm ring }}(t)$, 
constructed with the dominant black hole quasi-normal modes.
We observe that Eqs.~(\ref{Sk_341}) defining our semi-analytic puncture evolution 
display features and predictions qualitatively similar to what are observed by 
various Numerical Relativity
groups employing punctures to model black-holes. 

Efforts are being planned to compare the predictions of a suitably modified SAPE
with that arising from the modified EOB approach.
We have also initiated a program to include the spin effects into ${\cal H}_{\rm Sk}$.
Further,  due to the fact that  ${\cal H}_{\rm Sk}$ is exact
while describing the motion of a test black hole
in the Schwarzschild spacetime, a refined version of the SAPE should be 
of some interest to GW physicists, excited by the prospect of 
LISA observing Extreme/Intermediate Mass Ratio Inspirals.
 
\begin{acknowledgments}

It is our pleasure to thank Mark Hannam for discussions.
This work is supported in part by
the DFG (Deutsche Forschungsgemeinschaft) through SFB/TR7
``Gravitationswellenastronomie'' and
the DLR (Deutsches Zentrum f\"ur Luft- und Raumfahrt).

\end{acknowledgments}

\appendix

\section{ Formulae required to implement 
right hand sides of Eqs.~(\ref{Sk_341})}
\label{Sk:Appendix:A}
\begin{widetext}

   A close inspection of Eq.~(\ref{Eq22_Sk}) reveals that ${\cal H}_{\rm Sk}$ depends implicitly 
on $r, p_r $ and $j$. Therefore, evaluating right hand sides of Eqs.~(\ref{Sk_341}) in the SAPE
is rather involved. Using the theorem of implicit differentiation,
$ \partial {\cal H}_{\rm Sk} / \partial p_r$ may be computed using the following relation
\begin{align}
\label{dpsi2dpr}
\frac{\partial { \psi_2}}{\partial { p_r}}\biggl |_{r, j \,\,{\rm fixed}} &= 
\biggl \{ {\frac {\partial \psi_{2} }{\partial {p_r}}} 
+ \left( {\frac {\partial \psi_{{2}} }{\partial \psi_{{1}}} } 
 \right) \times
\left (
{\frac {\partial \psi_{{1}} }{\partial { p_r}}} 
\right ) \biggr \} \biggr / 
\biggl \{ 1- \left( {\frac {\partial \psi_{{1}}}{\partial \psi_{{2}}}
}
\right) \times 
\left ( {\frac {\partial \psi_{{2}} }{\partial \psi_{{1}}}}
\right ) \biggr \}
\,,
\end{align}
 and a similar relation holds for 
$\frac{\partial { \psi_1}}{\partial { p_r}} $.
We follow a similar scheme to compute the right hand sides of equations that define $d \phi/dt$ and $d p_r/dt$.

 During our matching to the QNMs, we require to compute  $^{(3)}{{\cal I}^{22}_{\rm plunge } }(t) $ and this demands
computation of $ d \omega/dt$ under the SAPE. From Eq.~(\ref{Sk_341b}), it is clear that 
$\dot \omega$ requires us to compute $d \psi_1/dt $ and $d \psi_2/dt$ and we use following relations:
\begin{subequations}
\label{dpsidt}
\begin{align}
\frac{d{ \psi_1}}{d { t}} &= 
\biggl \{  \left({\frac {\partial \psi_{{1}} }{\partial \psi_{{2}}}}
\right)
\kappa_{{2}}+\kappa_{{1}} \biggr \}
\biggr / \biggl \{ 
{1- \left( {\frac {\partial \psi_{{1}} }{\partial \psi_{{2}}}} 
 \right) \times \left (  {\frac {\partial  \psi_{{2}} }{\partial \psi_{{1}}}}
\right )
 }
\biggr \}
\,,\\
\frac{d{ \psi_2}}{d { t}} &=
\biggl \{  \left({\frac {\partial \psi_{{2}} }{\partial \psi_{{1}}}}
\right)
\kappa_{{1}}+\kappa_{{2}} \biggr \}
\biggr / \biggl \{
{1- \left( {\frac {\partial \psi_{{1}} }{\partial \psi_{{2}}}}
 \right) \times \left (  {\frac {\partial  \psi_{{2}} }{\partial \psi_{{1}}}}
\right )
 }
\biggr \} \,,
{\mbox\,\, {\rm where} \,} \\
\kappa_{1} &= \left( {\frac {\partial  \psi_{{1}} }{\partial r}} 
 \right) {\frac {dr}{dt}}
 + \left( {\frac {\partial \psi_{{1}}}{\partial { p_r}}}
  \right) {\frac {d p_r}{dt}} 
+ \left(
{\frac {\partial \psi_{{1}} }{\partial j}} 
 \right) {\frac {dj}{dt}} \,,\\
\kappa_{2} &= \left( {\frac {\partial  \psi_{{2}} }{\partial r}}
 \right) {\frac {dr}{dt}}
 + \left( {\frac {\partial \psi_{{2}}}{\partial { p_r}}}
  \right) {\frac {d p_r}{dt}}
+ \left(
{\frac {\partial \psi_{{2}} }{\partial j}}
 \right) {\frac {dj}{dt}} \,.
\end{align}
\end{subequations}

 We would like to emphasize Eqs.~(\ref{dpsidt}) are only required during our matching to the QNMs.


\newpage

\begin{figure}
\resizebox{15cm}{!}{\includegraphics{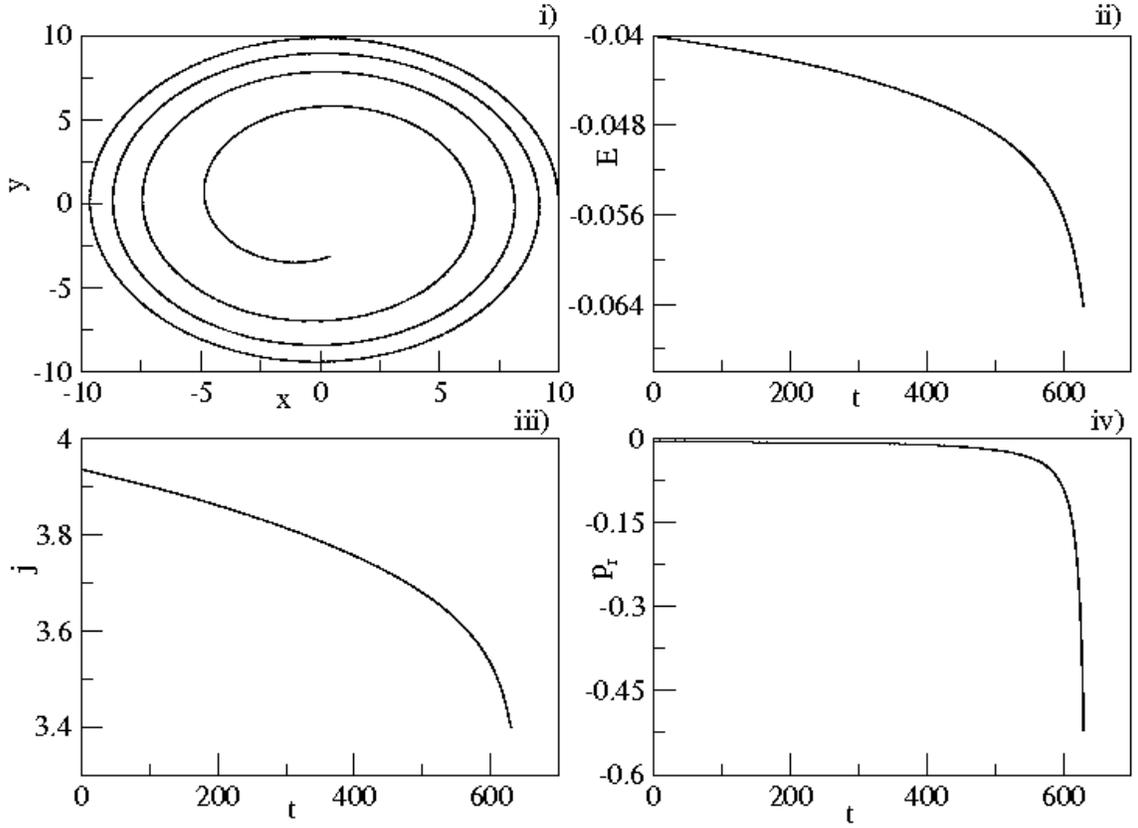}}
\caption { Plots depicting various facets of SAPE, defined by Eqs.~(\ref{Sk_341}), for a 
$\eta =0.25$ BBH having initial 
orbital separation $r=10$
[recall that we have dropped {\it hat symbol} appearing in the dimensionless variables
defined by Eq.~(\ref{Eq_21_Sk})].
In these plots, we terminate the SAPE when $\omega$ reaches its maximum value,
$\omega_{\rm mx} \sim 0.0896$.
The Panel~(i) provides a parametric plot of $x=r\, \cos \phi$ versus $y=r\, \sin \phi$, 
indicating that the SAPE occurs along quasi-circular orbits. 
The temporal evolutions for the dimensionless  binding energy, 
$ E \equiv {\cal H}_{\rm Sk}$,
and the orbital angular momentum $j$ are 
plotted in the panels~(ii) and (iii). The plot for $p_r(t)$ 
suggests that even during the dynamical plunge, {\it i.e.} in the neighborhood of $\omega_{\rm mx}$,  
the orbital motion is not that different from
a quasi-circular inspiral. 
}
\label{fig:xyEjpr}
\end{figure}
\newpage
\begin{figure}
\resizebox{15cm}{!}{\includegraphics{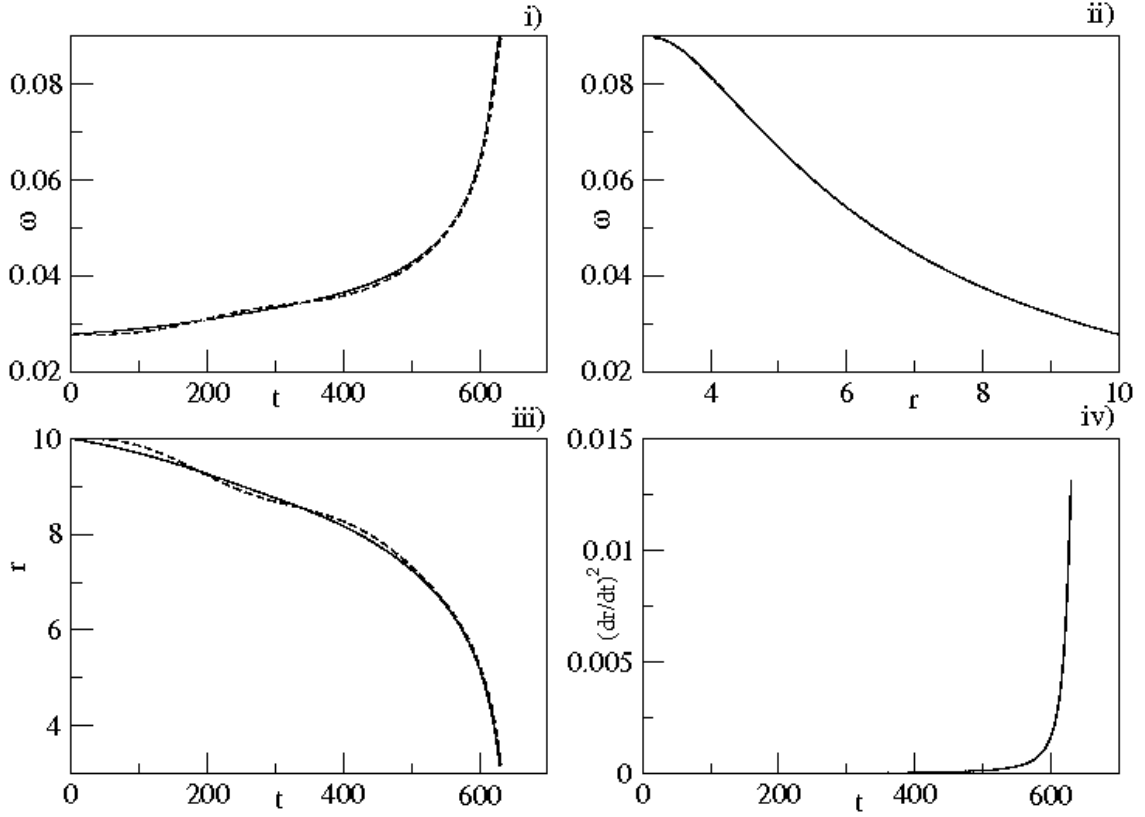}}
\caption { Plots providing $\omega(t), \omega(r(t)), r(t) $ and $\dot r(t)^2$ for 
BBH evolution under SAPE [the binary configuration is same as in Fig.~\ref{fig:xyEjpr}
and all quantities are dimensionless].   
The rapid changes in $\omega, r $ and $\dot r^2$ occurring during the dynamical plunge
is clearly visible and we terminate the SAPE when $\omega$ reaches its maximum value 
$\omega_{\rm mx}$.
Though $\dot r^2 = ( dr/dt)^2 $ remains small near $\omega_{\rm mx}$, its magnitude changes roughly 
fivefold during the plunge under SAPE. 
The black-hole binary evolution that employs  Eq.~(\ref{pr_ini}) 
for the initial value of $p_r$ is displayed with thick line,
while BBH evolution having $p_r =0 $ at the initial instant is displayed with dashed lines.
The spurious eccentricity in SAPE is suppressed by using at $t=0$ the expression for $p_r$, 
given by  Eq.~(\ref{pr_ini}), 
and this is also reported in Ref.~\cite{HHBGS}.
}
\label{fig:omgtomgrrtdrSt}
\end{figure}
\newpage

\begin{figure}
\resizebox{15cm}{!}{\includegraphics{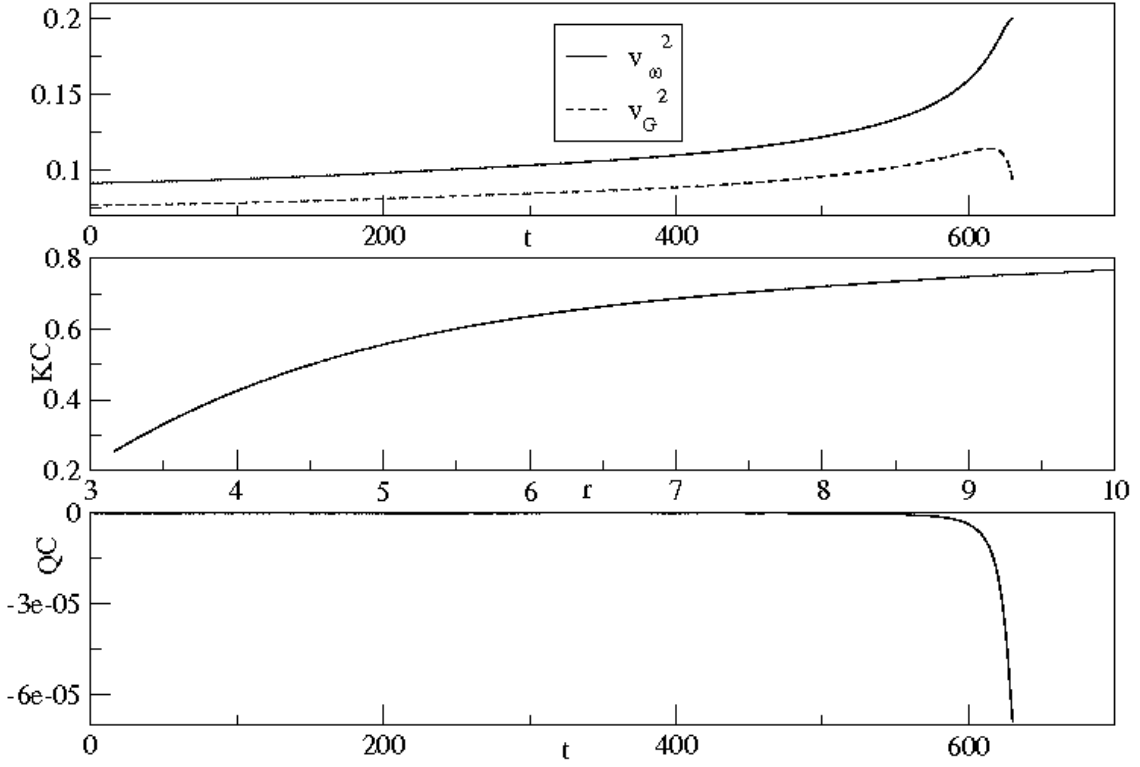}}
\caption { Plots displaying reactive evolutions of few interesting dynamical 
quantities and relations in the SAPE 
[the binary configuration is same as in Fig.~\ref{fig:xyEjpr}
and all quantities are dimensionless].
The top panel plots  $ v_{\omega}^2 =\omega^{2/3},
v_{\rm G}^2 = \dot r^2 + r^2\, \omega^2 $. The strong $r$
dependence of $v_{\rm G}^2$ makes it less attractive to characterize
the orbital velocity compared to $ v_{\omega}$. In the middle panel, we plot 
the `Kepler combination'(KC) $\omega^2\, r^3$  against $r$ and a gradual decrease   
is clearly visible throughout the inspiral. The quasi-circularity (QC) condition 
 $( d\,{E} - \bar \omega\, d\,j) $ is displayed in the bottom panel 
and its smallness justify the use of $ v_{\omega}$ in the 3.5 PN accurate expression for $dj/dt$.
}
\label{fig:tVS_KC_QC}
\end{figure}

\newpage
\begin{figure}
\resizebox{15cm}{!}{\includegraphics{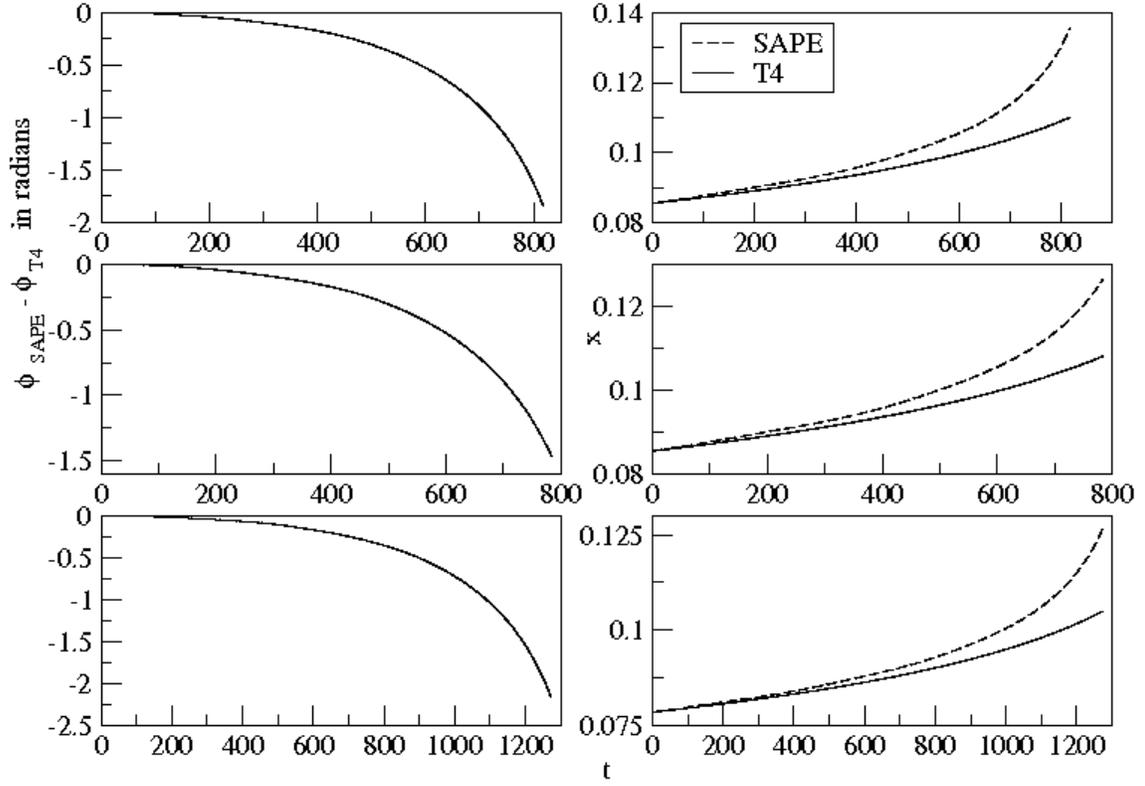}}
\caption {Differences in 
the orbital phase evolutions associated with the SAPE and TaylorT4 approximant at 3.5PN order
along with the associated $x \equiv \omega^{2/3}$ evolutions
[recall that we have dropped {\it hat symbol} appearing in the dimensionless variables
defined by Eq.~(\ref{Eq_21_Sk})].
From the top to bottom panels. the orbital frequency ranges are $[0.025-0.05], [0.025-0.045]$ 
and $[0.022-0.045]$ respectively.  
The fractional differences in $\phi$  are $\sim 7 \%, 5.8\%$ and $\sim 6.1 \% $
in the above three cases. A sharp increase  in  
$\phi_{\rm Sk}(t)- \phi_{\rm T4}(t)$ 
as we approach the plunge is also observed.
}
\label{fig:Sk_T4}
\end{figure}

\newpage
\begin{figure}
\resizebox{15cm}{!}{\includegraphics{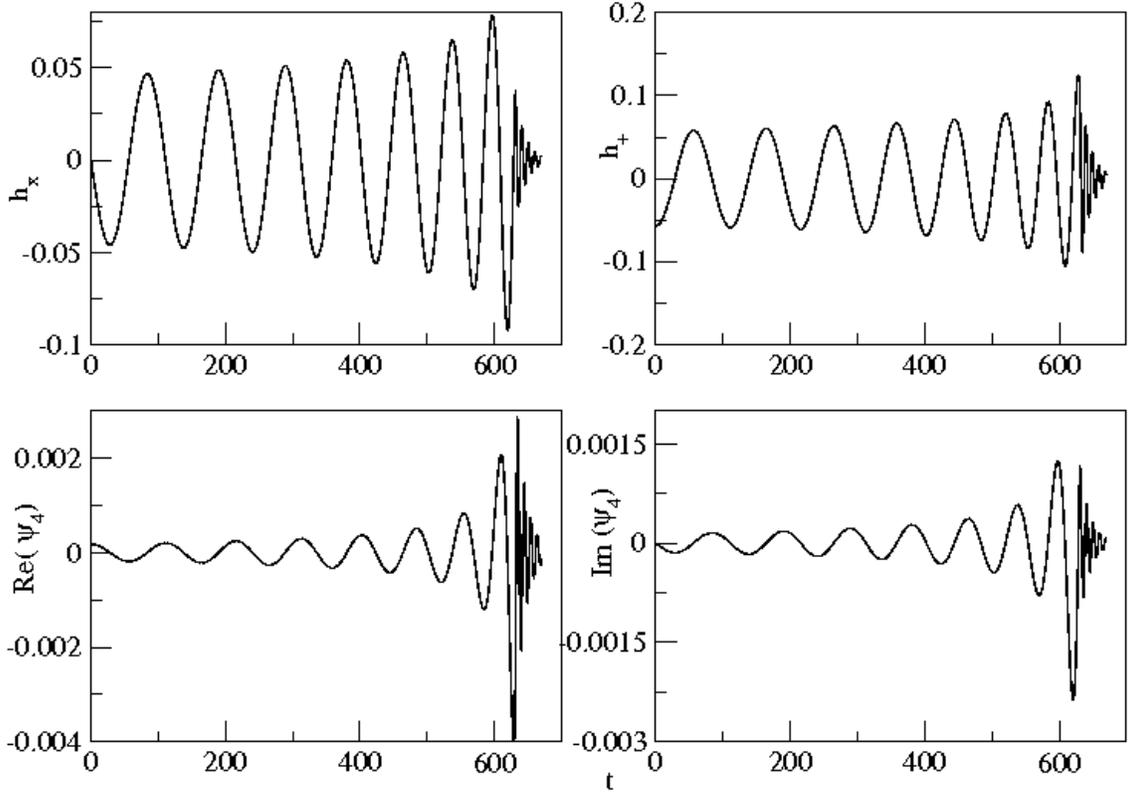}}
\caption {Plots showing temporal evolution of  
scaled $h_{\times,+} $, $\Re (\psi_4)$ and $\Im (\psi_4)$ for equal mass binary black-hole 
coalescence in SAPE. We scale out $G\,m/c^2\, r'$ from Eqs.~(\ref{EqSk_36}) and (\ref{EqSk_41}) and let $i = \pi/3$.  
The matching to the QNMs is based on Ref.~\cite{DG06} (see  Section~\ref{Skdyn_h}
and recall that $t$ is identical to $\hat t$, defined by Eq.~(\ref{Eq_21_Sk})).
}
\label{fig:thxppsi4RI}
\end{figure}

\end{widetext}

\end{document}